\documentclass{llncs}

\usepackage{balance}  % to better equalize the last page
\usepackage{graphics} % for EPS, load graphicx instead
\usepackage{times}    % comment if you want LaTeX's default font
\usepackage{url}      % llt: nicely formatted URLs

\usepackage{subfigure}
\usepackage{epstopdf}     
\usepackage{epsfig}
\usepackage{amsmath}

\newcommand{\specialcell}[2][c]{%for multi-line cells in tables
  \begin{tabular}[#1]{@{}c@{}}#2\end{tabular}}

\makeatletter
\def\url@leostyle{%
  \@ifundefined{selectfont}{\def\UrlFont{\sf}}{\def\UrlFont{\small\bf\ttfamily}}}
\makeatother
\def\pprw{8.5in}
\def\pprh{11in}

\usepackage[pdftex]{hyperref}
\hypersetup{
pdftitle={The Social Name-Letter Effect on Online Social Networks},
pdfauthor={LaTeX},
pdfkeywords={name-letter effect, Online Social Networks, same-name effect},
bookmarksnumbered,
pdfstartview={FitH},
colorlinks,
citecolor=black,
filecolor=black,
linkcolor=black,
urlcolor=black,
breaklinks=true,
}

\begin{document}

\title{The Social Name-Letter Effect \\ on Online Social Networks}

\author{Farshad Kooti\inst{1}\thanks{This work was done while the first two authors were interns at Qatar Computing Research Institute.} \and Gabriel Magno\inst{2}$^\star$ \and Ingmar Weber\inst{3}}
\institute{USC Information Sciences Institute, Marina del Rey, USA \and Universidade Federal de Minas Gerais, Belo Horizonte, Brazil \and Qatar Computing Research Institute, Doha, Qatar}

\maketitle

\vspace{-3mm}
\begin{abstract}
The Name-Letter Effect states that people have a preference for brands, places, and even jobs that start with the same letter as their own first name. So Sam might like Snickers and live in Seattle. We use social network data from Twitter and Google+ to replicate this effect in a new environment. We find limited to no support for the Name-Letter Effect on social networks. We do, however, find a very robust Same-Name Effect where, say, Michaels would be more likely to link to other Michaels than Johns. This effect persists when accounting for gender, nationality, race, and age. The fundamentals behind these effects have implications beyond psychology as understanding how a positive self-image is transferred to other entities is important in domains ranging from studying homophily to personalized advertising and to link formation in social networks. 
\end{abstract}

%\category{J.4}{Computer Applications}{Social and Behavioral Sciences}
%A category including the fourth, optional field follows...
%\category{H.1.2.a}{Models and Principles}{Human Factors}

%\keywords{Name-Letter Effect, Psychology Theory Validation, Twitter, Self Image}

%\textcolor{red}{The following section is mandatory for accepted papers, but not needed for submissions for the June 1 deadline.}
%\keywords{
%	Guides; instructions; author's kit; conference publications;
%	keywords should be separated by a semi-colon.
%}

%\category{H.5.m.}{Information Interfaces and Presentation (e.g. HCI)}{Miscellaneous}

%See: \url{http://www.acm.org/about/class/1998/}
%for more information and the full list of ACM classifiers and descriptors. 

%\textcolor{red}{The following section can be included for accepted papers, but is not needed for submissions for the June 1 deadline.}

%\terms{
%	Human Factors; Design; Measurement. 
%	If you choose more than one ACM General Term, separate the terms with a semi-colon.
%}

%See list of the limited ACM 16 terms in the instructions and additional information:
%\url{http://www.sheridanprinting.com/sigchi/generalterms.htm}.
%\textcolor{red}{Optional section to be included in your final version.}

% INTRODUCTION
\vspace{-2mm}
\section{Introduction}\label{sec:introduction}
\vspace{-2mm}
According to the \emph{Name-Letter Effect} (NLE), people have a preference for partners, brands, places, and even jobs that share the first letter with their own name. Correspondingly, a Sarah 
would be more likely to marry a Sam, go to Starbucks, move to San Francisco, and work in sales. This phenomena has been replicated in numerous settings \cite{nuttin85ejsp,nuttin87ejsp,hoorensetal90ejsp,hodsonolson05pspb,brendletal05jcr,pelham2005implicit,bekkers10nvsm} and is part of text books in psychology \cite{learytangney11}.
Some researchers have, however, questioned the validity or at least the generality of such studies \cite{simonsohn2011defense,simonsohn11apa,lebel2011sexy,mccullough2010baseball,dyjasetal12fp}. 
By its supporters, the NLE is usually attributed to ``implicit egotism'' \cite{pelham2005implicit} with people preferring situations that reflect themselves. %, akin to the phenomena of homophily.
%  FUND RAISING \cite{bekkers10nvsm}

%In the spirit of Computational Social Science, 
We turn to data from online social networks, Twitter and Google+, to see if the NLE can be replicated in a large online setting. Concretely, we seek evidence for or against the NLE in choosing social connections (Sarah following Sam) and in expressing brand interest (Peter following Pepsi). Our findings here are mixed and, depending on the exact setting, we find statistically significant evidence both for and against the NLE. % (in fact, even for a \emph{negative} NLE).

Extending the NLE and the idea of implicit egotism, we look for a \emph{Same-Name Effect} (SNE) where a Sarah is more likely to follow another Sarah and Tom Cruise is in particular popular among Toms. Here, we observe the presence of the SNE in different settings. We show that the SNE exists for both genders and in different countries. We also show that the SNE affects linking both to celebrities and to normal users and affects both strong and weak ties. Finally, we show that there is an anti-correlation between the number of friends and the extent of link preference bias caused by the SNE. % indicating that more selective users with fewer friends exhibit
%\todo[Ingmar]{@All: check that this holds for both networks.}
%\note[Gabriel]{@Ingmar: It holds for G+, still have to verify weak-strong ties}

%When studying the effects of NLE and SNE we make efforts to correct for as many circumstantial factors as possible. For example, we only look at users from the same country and look at the relative 

To the best of our knowledge, this is the first time that the Name-Letter Effect and the Same-Name Effect have been studied in an online setting. It is also the largest study of its kind with more than a million connections analyzed.
%\todo[Ingmar]{@Gabriel: how many edges did you effectively end up using? Do you know? Can we say > 10 million? Not the whole network, but only the same-name/same-letter ones we use.}
%\note[Gabriel]{@Ingmar: the G+ graph with US users have more than 20 million edges. Of course we do not use all the edges at the same time, since we'll filter for some names, but this is the potential number.}
 Our analysis quantifies a factor that affects link formation in online social networks. Understanding the processes governing which  links are established is crucial for areas such as information diffusion or link prediction. Moreover, the strength of the NLE or the SNE for an individual could be an estimate of the person's positive self-image. Understanding this could help in understanding homophily, and it could also be used in personalized advertising.

%\todo[Ingmar]{Add something on applications and understanding social link preferences in the age of an ever-more connected world.}
%\note[Farshad]{I've added a few sentences from the summary of the changes to here.}	
%\note[Ingmar]{This leaves a note. Can be removed by changing show to hide in the header.}
%\todo[Ingmar]{This leaves a TODO. Can be removed by changing show to hide in the header.}

% RELATED WORK
\vspace{-2mm}
\section{Related Work}\label{sec:related}
\vspace{-3mm}
%Discovery and cause
The NLE was first observed by Nuttin in 1985~\cite{nuttin85ejsp}. The effect was studied by asking volunteers to pick their favorite letter from pairs or triads of letters where only one of them belonged to the participant's first or last name. Nuttin showed that independent of visual, acoustical, semantic, and frequency characteristics, letters belonging to own first and last name are preferred over other letters. The most popular explanation for the NLE is ``implicit egotism''~\cite{pelham2002}. People have positive feelings about themselves and these feelings are associated implicitly to places, events, and objects related to the self~\cite{pelham2005implicit}.

%In different languages and cultures
Later, the presence of the effect was tested in different languages and cultures. It has been shown that the NLE exists in twelve European languages~\cite{nuttin87ejsp}. Also, Hoorens et al., showed that the NLE exists across languages, i.e., participants picked the letters in another language that were either visually or acoustically similar to the letters in their names in their own language~\cite{hoorensetal90ejsp}. %\todo[Ingmar]{@Farshad: Visually or acoustically similar letters?}

%In different areas
After the discovery of the NLE, many studies verified the existence of the effect in a wide range of decision making situations: People are disproportionately more likely to live in cities and take jobs that are similar to their name~\cite{pelham2002}. Also, brands that have the same initial as a person's name are preferred by that person~\cite{brendletal05jcr} and there is a higher chance of donation when the name of the solicitor is similar to the name of the contacted person~\cite{bekkers10nvsm}. Studies have even found that NLE affects marriage; people are more likely to marry a 
person with a similar name~\cite{jones2004love,simonsohn11apa}. On the other hand, the NLE was not observed in choosing favorite foods and animals~\cite{hodsonolson05pspb}.

%Questioning NLE
Besides many studies providing evidence for an NLE, there have also been papers questioning the presence of the effect in different areas or the reason for the effect. E.g., in~\cite{mccullough2010baseball}, the authors show that a wrong statistical test was conducted in an earlier work on verifying the existence of the NLE in the initial of a baseball player and number of strike outs by him. Also, other works had shown different biases that might create the same results as a NLE~\cite{simonsohn2011defense,simonsohn11apa,lebel2011sexy}. For example, in the study that showed people are more likely to live in cities with the same initial as theirs, one explanation might be that people in those cities named their babies with such names. Although there are some papers challenging the existence of NLE, the critics are usually concerned about the way a particular study was done, and the main effect is still generally accepted.

%Our work
%In the current work, we test the NLE in a new environment, namely in an online social network. We investigate the NLE in preference of linking both to brands and other users. Moreover, we introduce a more restricted effect called same-name effect and study its existence in various setups.
\if 0
We are aware that a preference to form links to users with similar names could also be due to homophily. E.g., different social circles might follow different naming conventions for their children.
 But we try to rule out as many confounding factors as possible. To this end, in our analyses
we account for gender, nationality, age, and race of users by considering their first and last name and the location they provided in their profile. We conducted our tests in a way that 
the analyzed sample is homogenized with respect to the main characteristics of a person and this will mitigate the extent to which homophily exists between users.
\fi
%\todo[Ingmar]{@Farshad: Here might also want to add that, being aware of such criticism, we try to rule out as many other confounding [use this term] factors as possible.}
%\note[Farshad]{I've added the last para. This is what you meant, right? If I got it right, doesn't this explanation fit better in the introduction rather than related work?}

%\cite{bertrand2003emily} The study on job application and name, might be added later
%\cite{dyjasetal12fp}, 

% THE NAME-LETTER EFFECT ON TWITTER
\vspace{-1mm}
\section{The Name-Letter Effect on OSNs}\label{sec:nle}
\vspace{-2mm}
%As mentioned earlier, the NLE has been observed in various settings. 
In this section, we first test the generality of the NLE on Twitter and Google+ in different domains, such as preferred brands, celebrities, and news media. Then, we investigate the NLE in the social context. Concretely, do users follow other users with the same initial disproportionately more than users with a different initial? Here we use the term ``follow'' to refer both to Twitter following and to Google+ ``has added to a circle''. In both case, the acting user expresses an interest in the updates of the user acted upon.

\vspace{-1mm}
\subsection{Data Description}
\vspace{-2mm}
{\bf Twitter:} Most of the analyses in this work is done on a large Twitter social network gathered in~\cite{icwsm10cha}. The network contains all the 52 million users who joined Twitter by September 2009 and all the 1.9 billion links among them. We also used users' location information from~\cite{kulshrestha12icwsma}, which uses both location and time zone fields for inferring a user's country.

%\note[Gabriel]{Not sure were to put description of the Google+ dataset}
%\todo[Ingmar]{@Gabriel: are some of the details also in the cited paper? If yes, can we just cite that paper and shorten it to a similar length of the Twitter description?}

{\bf Google+:} The Google+ dataset was created by collecting public information available in user profiles in the network. %We inspected the \textit{robots.txt} file and followed the sitemap to retrieve the URLs of Google+ profiles.
%Since we retrieved the complete list of profiles provided by Google+, we believe we have all the users with public profiles in Google+ by the time of the second data collection.
The data collection ran from March 23rd of 2012 until June 1st of 2012. 
%When inspecting the sitemap we found 193,661,503 user IDs. 
In total we were able to retrieve information from 160 million profiles
%, since some IDs were deleted or we were not able to parse their information
. With the social links of the users, we have constructed a directed graph that has 61 million
nodes and 1 billion edges. 
%To identify an user's country, we extracted the geographic coordinates of the last location present on the~\textit{Places lived} field and identified the corresponding country. We were able to identify the country of {22,578,898} users.
Details of the Google+ platform and a data characterization of an early version of the dataset is discussed in a previous work~\cite{magno2012}.

\vspace{-1mm}
\subsection{NLE and Brand Preference}
\vspace{-2mm}
%Briefly mention brand congruence in social networks \cite{singlaweber11tweb}
%\todo[Ingmar]{@Gabriel: Did you do the same? So can we just replace ``Twitter'' by ``Twitter and Google+''?}
%\note[Gabriel]{Edited to take Google+ into account}
For testing the NLE on Twitter and Google+, we considered a variety of domains and 
we picked a pair of popular Twitter and Google+ accounts from each domain. Then, we gathered 
all the followers of each account as of May 2013 (or a large 1 million uniform, random sample of them) in Twitter,
and all the followers of each account in Google+ as in the time of the data collection (2012).

We examine the brand NLE by performing the Pearson's chi-squared test of independence. We do this by counting the followers of each account who have the same initial to see if there are disproportionately many followers for the brands and users with the 
same initial. For each pair of brands, we create a 
2 $\times$ 2 table showing the number of followers for each account whose initial is the same
as initial of either of brands. Since both the popularity of the brands and
the frequency of name initials are not necessarily the same across the world, in all the analyses in this
section we only consider followers in the US. To filter the users in Twitter we used the location field
from the users' accounts and only picked users who had one of the top 20 most populated
US cities, {\it United States}, or {\it USA} in their location field. 
The location filter in Google+ was done by extracting the geographic coordinates of the last location
present on the~\textit{Places lived} field, picking only the users from USA.

Table~\ref{table:brandNLEillust} shows an illustrating example of the 2 $\times$ 2 tables. 
$A$ represents the number of users who follow Brand 1 and have the same initial as Brand 1. 
Similarly, $D$ is the number of followers of Brand 2 who have the same initial as Brand 2.
For testing the NLE, 
first, we calculate the expected values for the cells that the initial of the followers matches
the brand's initial (here $A$ and $D$). The expected value, is the value that the fields
would have if, given the total values, the followers were split uniformly and without any preference. Here the
expected value of $A$ would be $\frac {(A + C) * (A + B)} {A + B + C + D}$ and the expected
value of $D$ would be $\frac {(B + D) * (C + D)} {A + B + C + D}$. Then, expected values
smaller than the observed values for $A$ and $D$ indicate the existence of the NLE. 

\if 0
%\vspace*{-2mm}
\begin{table} [b!]
\parbox{.45\linewidth}{
\centering
\begin {tabular} { c  c  c  c }
\hline
 & Brand 1 & Brand 2 & Total \\
\hline
Brand 1 initial & $A$ & $B$ & $A + B$\\
Brand 2 initial & $C$ & $D$ & $C + D$\\
\hline
Total & $A + C$ & $ B + D$ & \\
\hline
\end{tabular}
}
\hfill
\parbox{.45\linewidth}{
\centering
\begin {tabular} { c  c  c  c }
\hline
 & Jim Carrey & Tom Cruise & Total \\
\hline
Initial J & {\bf 3,266 (+2\%) }& 1,663 (-4\%) & 4,929\\
Initial T & 2,141 (-3\%) & {\bf 1,270 (+6\%)}& 3,411\\
\hline
Total & 5,407 & 2,933 & \\
\hline
\end{tabular}
}
\caption{The left table illustrates the test of NLE. If link formation is independent of the initials of the brands, the observed value would be close to the expected value for $A$, namely, $\frac {(A + C) * (A + B)} {A + B + C + D}$. Larger than expected observed values for $A$ and $D$ indicate the existence of the NLE. The right table shows a real-world numerical example and the effect sizes are show in parentheses. The observed counts for the diagonal of the matrix are larger than the expected counts for the followers with the same initial, validating the NLE.}
\label{table:brandNLEillust}
\end{table}
\fi

\begin{table}[b!]
\begin {center}
{
\begin {tabular} { c  c  c  c }
\hline
 & Brand 1 & Brand 2 & Total \\
\hline
Brand 1 initial & $A$ & $B$ & $A + B$\\
Brand 2 initial & $C$ & $D$ & $C + D$\\
\hline
Total & $A + C$ & $ B + D$ & \\
\hline
\end{tabular}
}
\end{center}
\caption{Illustration of testing the NLE. If link formation is independent of the
initials of the brands, the observed value would be close to the expected value for $A$, namely, $\frac {(A + C) * (A + B)} {A + B + C + D}$.
Larger than expected observed values for $A$ and $D$ indicate the existence of the NLE.}
\label{table:brandNLEillust}
\end{table}

\if 0
As a real numerical example, Table~\ref{table:brandNLEillust} shows the number of Twitter followers of 
two actors, Jim Carrey and Tom Cruise, and their followers in the US with a first name initial of 
either J or T. The expected values are calculated with the aforementioned equation. 
The observed value of 3,266 vs.\ the expected value of 3,195 shows that
users with the initial J have a 2\% increase in preference to follow Jim Carrey rather than Tom Cruise. 
A similar inclination is observed in users with the initial T towards following Tom Cruise.
We also calculate the $\chi^2(1)$ to evaluate the statistical significance, which is the confidence in refuting the null hypothesis that the rows are independent.
\fi

Table~\ref{table:brandNLE} shows all the considered Twitter and Google+ accounts and whether a
significant NLE exists or not. We picked these pairs mainly because these accounts have high number of followers. Moreover, the pairs presented here and in the rest of the paper are all the pairs that we did the analysis on, and we are not ``cherry-picking'' the results.
Out of the eight considered domains  in Twitter, shown in Table~\ref{table:brandNLE}, only three of them show
a statistically significant NLE, three cases imply NLE but the results are not statistically significant, and the
remaining two pairs exhibit a negative NLE. The results suggest that the NLE exists only in
some special cases and it is not a generalizable concept for following brands on Twitter.
This analysis was done by considering the first name of the user. We repeated the analysis
using the Twitter handle (i.e., screen name) of the users. For 61\% of the users the initial of the
actual name matches the initial of their Twitter handle. Due to this high
overlap, testing the NLE by using the handles yields very similar results to using their declared names: in only two cases
the results are statistically significant, for the game consoles and the actors, and in
these cases the effect is much smaller than the NLE with actual names (3.5\% and 1\% respectively).
In Google+, none of the three pairs of brands/celebrities had statistically significant results, 
with two of the pairs exhibiting low positive and one negative NLE.

%\todo[Ingmar]{In only two cases out of how many? For which setting? The results for which test are significant?}

\if 0
\begin{table}[t!]
\begin {center}
{
\begin {tabular} { c  c  c  c }
\hline
 & Jim Carrey & Tom Cruise & Total \\
\hline
Initial J & {\bf 3,266 (+2\%) }& 1,663 (-4\%) & 4,929\\
Initial T & 2,141 (-3\%) & {\bf 1,270 (+6\%)}& 3,411\\
\hline
Total & 5,407 & 2,933 & \\
\hline
\end{tabular}
}
\end{center}
\caption{An example of a brand NLE on Twitter. The effect sizes are show in
parentheses. The observed counts for the diagonal of the matrix are larger than the expected counts for the followers with the same initial,
validating the NLE. Also, $\chi^2(1) = 10.8$ showing
the significance at a $p-value < 0.001$.}
\label{table:brandNLEexample}
\end{table}
\fi

\vspace*{-3mm}
\begin{table} [b!]
\parbox{.45\linewidth}{
\centering
\texttt{Twitter}\\
\begin {tabular} { c  c  c  c  }
\hline
Account 1          & Account 2         & NLE       & p-value    \\
\hline
Sega               & Nintendo          & 9\%       & $< 0.001$  \\
Jim Carrey         & Tom Cruise        & 4\%       & $< 0.001$  \\
Firefox            & Internet Explorer & 5\%       & $< 0.1$    \\
Canon              & Nikon             & 5\%       & ---        \\
Puma               & Adidas            & 0.9\%     & ---        \\
CNN                & New York Times    & 0.4\%     & ---        \\
Nokia              & Samsung           & -1.3\%    & ---        \\
Pepsi              & Coca-Cola         & -1.7\%    & ---        \\
\hline
\end{tabular}
}
\hfill
\parbox{.45\linewidth}{
\centering
\texttt{Google+}\\
\begin {tabular} { c  c  c  c }
\hline
Account 1          & Account 2         & NLE    & p-value \\
\hline
Sergey Brin        & Larry Page        & 1\%    & ---     \\
Nokia              & Samsung           & -16\%  & ---     \\
Pepsi              & Coca-Cola         & 1\%    & ---     \\
\hline
\end{tabular}
}
\caption{The Twitter and Google+ accounts considered for the brand NLE and the average percentage of preference
for the brands with the same initial. There is no significant NLE for most of the brands.}
\label{table:brandNLE}
\end{table}
\vspace*{-1mm}

% \todo[Ingmar]{Not sure, but would it not be better to use the median and not the average everywhere? Only if that's feasible though and does not take too long to compute.}

\subsection{NLE and Social Link Preference}
\vspace*{-1mm}
In this section, we test the NLE in the context of  friend link preference. This means that we check if users prefer to establish links to other users with the same initial. To have two sets of users with the same initials for testing the NLE on link 
preference, we first picked the four most popular names on Twitter that have pairs of same initials: ``Michael'', ``Matthew'', 
``Jason'', and ``James''. %Similarly, for Google+ we chose: ``Michael'', ``Mark'', ``James'' and ``John''.
 Since these names are used in many countries, considering
all users might falsely show the NLE: say ``Michael'' and ``Matthew'' are 
popular in a particular country, but not in others, in this case there will be lots of links 
from ``Matthew'' to ``Michael'', but not to ``Jason''. This could create an apparent NLE in
the results, that might not actually exist, or at least not due to implicit egotism. To overcome this issue, we limited ourselves 
to users in the US.

Table~\ref{table:NLELink} shows the results of the number of times ``Matthews'' and
``Jameses'' follow ``Michaels'' and ``Jasons'' for Twitter. Surprisingly, the results show a
slight, statistically significant\footnote{In this work, we consider $p-value < 0.001$ as statistically significant, unless explicitly specified.} \emph{negative} NLE ($\chi^2(1) = 15.58$).
This analyses was repeated with a pair of female names (``Melissa'' and ``Jennifer'') following 
a pair of male names (``Michael'' and ``Jason'') and vice versa. Again in both cases a 
negative NLE existed, but this time not statistically significant. The results clearly show 
that the NLE does not exist for general social link preference. The same analyses were done for the Google+ dataset, using the two most popular pairs of same initials: ``Michael'', ``John'', ``Marks'' and ``James''.  Again, there was a negative NLE, but not a statistically significant one.

%\note[Gabriel]{For G+, if I consider the same list of names as in Twitter, there is a significant POSITIVE effect size of 6\%, contradicting Twitter. I will check the results}
%\note[Ingmar]{@Gabriel: That is interesting. But let's follow the same methodology of using the top-k names. So no need for the same names.}

\begin{table} [t!]
\parbox{.45\linewidth}{
\centering
\texttt{Twitter}\\
\begin {tabular} { c  c  c  c }
\hline
         & Michael            & Jason              & Total \\
\hline
Matthew  & {\bf 6,455 (-2\%)} & 4,285 (+4\%)       & 10,740\\
James    & 12,016 (+1\%)      & {\bf 7,236 (-2\%)} & 19,252\\
\hline
Total    & 18,471             & 11,521             & \\
\hline
\end{tabular}
}
\hfill
\parbox{.45\linewidth}{
\centering
\texttt{Google+}\\
\begin {tabular} { c  c  c  c }
\hline
         & Mark               & James              & Total \\
\hline
Michael  & {\bf 3,605 (0\%)} & 1,829              & 5,434 \\
John     & 3,213              & {\bf 1,598 (-1\%)} & 4,811 \\
\hline
Total    & 6,818              & 3,427              & \\
\hline
\end{tabular}
}
\caption{Results of the NLE on link preference. Effect sizes are shown in the parentheses.
In Twitter, users with same initials have negative effect size, contradicting the NLE ($p-values < 0.001$). Google+ results were not statistically significant. }
\label{table:NLELink}
\vspace{-3mm}
\vspace*{-5mm}
\end{table}

\vspace*{-2mm}
\if 0
\subsection{NLE and Word Usage}
On Twitter, users can provide a brief description or ``bio'' of themselves, which will be shown in their
profile under their name. We tried to see if the users show any NLE with respect to their general
word choice, i.e. do users disproportionately use more of the words that start with their initial or not.
To this end, we created a 26 $\times$ 26 table of the counts of number of words with different 
initials used by the users. We eliminated the name of the users from the bio, to avoid the potential
bias. We gathered 900K random users who have provided their bio and tested to see if the effect
exists. Out of the 325 pairs of possible combinations, 305 show statistically significant results,
and out of those 204 show positive NLE and the remaining 101 demonstrate negative NLE. So, the NLE
is not a generalizable concept for words choice.
\fi

\vspace{-1mm}
\subsection{NLE and Location, Job, and Hobbies}
\vspace{-2mm}
%\note[Gabriel]{Edited to include Google+ results. The results were almost identical to Twitter}

Earlier studies have shown people prefer to live in the cities with the same initials and also choose occupations that have the same initial as their name~\cite{pelham2002}. We tried to replicate these findings using our data. For Twitter we gathered the profile information of more than 4 million random users and used their location field to see the effect of NLE in the city that people choose to live. For Google+ we retrieved the city from the ``Places lived'' field. We tested the NLE for the top ten largest city  in the US\footnote{\url{http://en.wikipedia.org/wiki/List_of_United_States_cities_by_population}}. The ten largest cities in the US have seven unique initial letters, which leads to 21 (seven choose two), pairs of letters for checking NLE. In Twitter, out of the 21 pairs, 8 pairs show statistically significant results, with 6 of them showing positive NLE. In Google+, 7 pairs were statistically significant, with 6 of them showing positive NLE.

Similarly for the occupations, we consider the following jobs: engineer, cashier, waiter(ess), teacher, and nurse. In Twitter we search the users' bios for the corresponding strings. The ``bio'' is the field in the profile that users introduce themselves in and they often include their occupation. In Google+, we examine the ``Occupation'' field, and looked for the same set of strings. Both in Twitter and Google+ we find only one statistically significant result out of the ten (five choose two) possible pairs of letters, and this single statistically significant pair has negative NLE.

We also test the NLE for the hobbies of the users. More specifically, we look for popular sports in the bio of the users in Twitter and in the ``Introduction'' field of Google+. We consider football, basketball, baseball, lacrosse, soccer, volleyball, tennis, and hockey. We test the NLE again for the all 21 possible pairs of initials of the sports and the names. Only four of the pairs show a statistically significant result, with only one positive NLE in Twitter and two positive NLE in Google+.

Overall, our findings therefore question the existence or at least the general scope and strength of the NLE as we failed to replicate earlier claims in this new setting.

% THE SNE ON TWITTER

\section{The Same-Name Effect on OSNs}\label{sec:sne}
\vspace{-2mm}
%In the previous section, we showed that the NLE does not exist in most of the different domains and does not seem to influence the preference for link creation.
 In this section, we test another effect in 
link creation preference in a more restricted case where both users have the exact same first name, rather than
just the same initial. Since all letters of the users' names are involved, this effect should be stronger than the NLE.
We call this effect {\it same-name effect} (SNE). In other words, are Michaels
disproportionately more likely to follow other Michaels compared to other users? 
A similar idea was tested in an earlier study, where it was shown that people are more 
likely to marry others with the same \emph{last} name~\cite{jones2004love}. Here, we analyze linking 
between users with the same name and show that
there is a strong SNE that is robust to many variations.

%\todo[Ingmar]{@Gabriel+Farshad: The p-value above, which data set does it refer to?}
%\note[Gabriel]{@Ingmar: it is the same for both datasets. Should we mention it separated?}

%\emph[Ingmar]{Do we also have the summary of all pairwise effect sizes? Probably as the median instead of average. Just to have things comparable. Not crucial but cleaner I think.}
%\vspace{-3mm}
\if 0
\vspace{-1mm}
\subsection{SNE and Gender, Ethnicity, and Age}
\vspace{-2mm}
\fi
First, we test the SNE by considering the gender of users as the first name typically identifies the gender. Since men (women) might be more
likely to follow other men (women) \cite{magnoweber14socinfo}, considering both groups together might cause a false 
indication of a SNE. So, we perform the SNE test within each gender.
Also, as mentioned earlier, having users from different countries might introduce a bias
in the results, so again we are considering only users in the US.

We pick the five most popular male names on Twitter among users from the US: ``Michael'', ``John'', 
``David'', ``Chris'', and ``Brian''. Then, we count the number of times each of the users with these
names follows other users with these names. Table~\ref{table:SNE} (appendix) shows the resulting 
5 $\times$ 5 table and the effect sizes of 4-13\% on Twitter. We calculate the effect size of each name as the 
average of pairwise preference of that name over other names in the table. This same analysis is repeated in Google+, and the results are the same: male users significantly preferred to follow other users with the same name 7-30\% more than expected. 
We also tested the SNE with the five most popular female names in the US on Twitter and Google+: 
``Jennifer'', ``Jessica'', ``Ashley'', ``Sarah'', and ``Amanda''. The results were similar
to the previous case and even stronger: female users significantly preferred to follow others with the 
same name 30-45\% more than expected in Twitter, and 10-29\% in Google+.

%\todo[Ingmar]{@F+G: The p-value here is again for G+? Just to know. Do we have it for T as well?}
%\note[Gabriel]{@Ingmar: it is the same for both datasets. Should we mention it separated?}

%\emph[Ingmar]{Again, best to use median everywhere I think, if possible.}
An alternative explanation for the observed preference could be the fact that different 
first names are popular in different ethnicities and races. To address this concern, we 
repeated the analysis for all male first names in the US with more than 10,000 users (56 names in Twitter, 58 names in Google+). 
We tested the SNE pairwise for these names and the SNE existed for all 1,540 pairs of names with an
average effect size of 19\% in Twitter, and for all 1,653 pairs of names with an average effect size of 28\% in Google+. The fact that the SNE exists for all of the pairs suggests that
the preference is not just because of homophily because for at least some cases the names
would be associated with the same particular race or ethnicity.

%\note[Gabriel]{@Farshad: are the ``top 50 popular first names'' the Top 50 overall for US, or the top 50 of each race? For Google+ I considered the top 50 overall}
%\note[Farshad]{@Gabriel: I'm using top 50 in each group, since I thought top 50 will be dominated by the US, and that's not desired. I guess the results wouldn't be that different. We should just fix it for the camera-ready/later submission.}

Moreover, we used last names as a proxy of the ethnicity. 
We used 1990 census data to gather last names that are prominent for only one race in the US.\footnote{\url{http://names.mongabay.com/}} We gathered
the top 1,000 last names in each of the five races of white, African-American, Asian, Hispanic,
and native American natives.\footnote{Note that in later census the race/ethnicity has been treated differently and that ``Hispanic'' can now be of any race according to the census terminology.} For each race
we considered only the last names that are in the top 500 of a particular race and do not occur among the
 top 1,000 names for any of the other race's lists. Then we tested the SNE within each race for the pairwise combination of the top 50 popular first names, 1,225 pairs, though not all of these 50 first names were found for all of the five races.  Table~\ref{table:race} shows that for all five races a strong and consistent SNE exists in Twitter. In Google+, most of the results were not statistically significant, although implying positive SNE. 

To account for age, we use offline data from social security statistics\footnote{\url{http://www.ssa.gov/oact/babynames/top5names.html}}. We focus on the common ages of 20-30 years old  on Twitter at the time the data was collected (2009), which corresponds to users born between 1979-1989. We use the records of social security to gather the most popular boy baby names during the mentioned years. Then, we pick all the names that were in the top five at least once: ``Michael'',  ``Jason'', ``Christopher'', ``Matthew'', ``David'', ``James'', and ``Daniel''. We conduct a similar analysis to the previous section on these names. A statistically significant SNE again existed with 12-17\% preference in Twitter, and 5-23\% statistically significant preference in Google+. We also try the same experiment with the most popular girl baby names during 1979-1989. Again, a significant SNE is observable with a 16-24\% preference in Twitter and 10-106\% preference in Google+.

Finally, to see if the SNE  exists in different languages and cultures, we picked three countries
with different languages: Brazil, Germany, and Egypt. Then, we picked the most
popular names in each of those countries and tested the SNE. We found that a statistically significant SNE exists in all three countries, 
both for Twitter and Google+. The effect sizes for Brazilian users range   from 13-22\% in Twitter and from 16-22\% in Google+.
Similarly, in Germany and Egypt users significantly preferred to follow other users with the same name (6 - 101\%).

\vspace{-1mm}

\vspace{-2mm}
\if 0
\section{Implications}\label{sec:implication}
\vspace{-2mm}
%Talk about potential explanations, possible weaknesses and extensions
%\note[Ingmar]{I'll add points here later today.}

In this study, we quantified a factor that affects link formation in online social networks. Understanding the way
in which links are established is necessary for various areas such as information diffusion and link prediction.

The Name-Letter Effect is a well-studied subject in psychology
with dozens of studies dedicated to it as the NLE affects several aspects of people's offline life.
These aspects include important decisions such as cities to live in, jobs to choose, and choosing a life partner. 
One study has even shown that the NLE can be 
exploited to increase the chance of donation in fundraising~\cite{bekkers10nvsm}.
Similarly, it is easy
to imagine similar situations in the online world. If a strong NLE had existed on establishing
social links, one direct application could have been the recommendation of friends, celebrities, and 
brands to follow. The same idea applies to the SNE: recommendation
algorithms can assign higher scores to friends and celebrities that have the same name as the user, or this could be one element in the assignment of teams.
\fi
%INSERT CITATION FOR THE FUNDRAISING PAPER.

\vspace{-2mm}
\section{Discussion}\label{sec:discussion}
\vspace{-2mm}
We have focused on testing and observing the NLE and the SNE rather than on explaining them. 
When using implicit egotism as an explanation the crucial assumption is that users are free to 
choose the brands they like or the members of their social network. This basic assumption is 
arguably flawed as people can only connect to people (or brands) they know. But as the distribution 
of names is not homogeneous across all parts of society this creates implicit selection biases. 
For example the name Emma was very popular for girls born during 2002-2012\footnote{\url{http://www.ssa.gov/oact/babynames/top5names.html}} but less popular earlier 
which, in turn, means that an Emma would be more likely to go to school with another Emma and hence have a chance
 to connect. Similarly, the name DeShawn is popular among African Americans~\cite{fryer2004causes} which 
 means a DeShawn growing up in a predominantly black neighborhood would again have a higher than 
 expected chance of connecting to another DeShawn. 
%Similar effects are induced by names popular  among particular religious groups, among particular geographic regions, or among particular classes  of income and education.
 In fact, previous research has shown that mere familiarity with a name 
 correlates with likeability \cite{colman1980psychological,hargreaves1983attractiveness}.

%\todo[Ingmar]{This is only relevant if the name was *not* popular during other time periods. If it is always popular then this is an empty statement.}
%\note[Farshad]{I've checked the list and Emma is on top 5 only after 2002. Added ``'but less popular earlier'}

% \note[Gabriel]{The Google+ dataset was also collected in an early stage: launched 3rd quarter of 2011, collected 2nd quarter of 2012}

%Though it is hard to correct for all such factors, 
We tried to avoid obvious pitfalls,  such as selecting names associated with a particular demographic groups, and we  looked at names that were popular during a certain period. Additionally, the fact that for 
testing the NLE and the SNE on link preference we only used the network of early adopters of Twitter (up to September 2009) and Google+ (less than a year after the launch) helps to further homogenize the user set across age and income. Also, we have used users' last names to test the SNE within one race. Still, naming conventions within a family, where family members are given the same first name, could explain part of the observed the SNE. 

\if 0
Concerning brands, some effects could be induced by the fact that different brands enjoy different popularity with different groups in society. For example, if Pepsi is more popular with the  ``New Generation''\footnote{\url{http://en.wikipedia.org/wiki/Pepsi_Generation}} then this  would correspond to a difference in the distribution of names of its customers. Even stronger  effects relate to different market shares in different countries if international customers were to be considered.
\fi

It is also not clear what fraction of users use their real name in online social networks. We believe this is the case for the majority of the users, especially for Google+, since Google explicitly asked users to use their real name and banned the accounts of users with fake names\footnote{\url{http://gawker.com/5824622/names-banned-by-google-plus}}. There might be much less use of real names on Twitter, but the fact that our findings for Twitter and Google+ are very consistent suggests that there is no dramatic difference between Twitter and Google+ in the way people chose their name. And even if the majority of the names are not real, we still found the SNE, which might have a different explanation than the implicit egotism. Also, note that for testing SNE, we tested the effect on common English names, so we are not analyzing users completely fake names like ``cowboy''.

\vspace{-2.3mm}
\section{Conclusions}\label{sec:conclusions}
\vspace{-2.3mm}
The Name-Letter Effect (NLE) states that people prefer the letters in their own names
over other letters. We investigated the existence of the NLE in the context 
of Twitter and Google+. Our findings question at least the generality of the NLE. %First, we tested the NLE's influence on linking to brands to see if users follow brands with the same initial disproportionately more than other brands. Though we do find a weak but statistically significant NLE for some brands, we find no evidence for others.
Going beyond the NLE, we analyzed users' linking behavior for a same-name effect (SNE),  where instead of comparing the initials we compared the whole name. In this stronger version, we observe a robust effect, even when accounting for gender, age, race, and location.

Besides the psychological aspects of NLE, there are some real-world implications. E.g., one study has showed that using NLE can increase the chance of donation made by people~\cite{bekkers10nvsm}. In recent years, the Coca-Cola {\it share a coke\footnote{\url{http://www.coca-cola.co.uk/faq/products/share-a-coke.html}}} campaign has proven to be very successful by increasing sales\footnote{3MM P4W Consumption Oct-Dec 2011 B3 Survey Australia}.
\bibliographystyle{splncs}
\bibliography{NameLetterEffectOnTwitter}

\appendix 

\section{Appendix}
\begin{table*}[tbh!]
\begin {center}
\texttt{Twitter}\\
{
\begin {tabular} { c c c c c c c}
\hline
&Michael & John & David & Chris & Brian & Total \\
\hline
Michael& \bf{28,587 (+5\%)}	&	36,590	&	29,051	&	25,928	&	15,093	&	135,249\\
John & 28,393	&	\bf{42,417 (+4\%)}	&	31,540	&	27,823	&	16,906	&	147,079\\
David & 24,303	&	33,713	&	\bf{29,388 (+5\%)}	&	24,441	&	14,255	&	126,100\\
Chris & 22,632	&	31,383	&	25,107	&	\bf{25,999 (+6\%)}	&	14,089	&	119,210\\
Brian & 15,394	&	20,974	&	16,676	&	15,636	&	\bf{11,383 (+13\%)}	&	80,063\\
\hline
Total & 119,309	&	165,077	&	131,762	&	119,827	&	71,726	&\\
\hline
\end{tabular}
}
\end{center}

\begin {center}
\texttt{Google+}\\ 
{
\begin {tabular} { c c c c c c c }
\hline
 & Michael & David & John & Chris & James & Total \\ 
\hline
Michael & {\bf5,949 (+7\%)} & 4,492 & 4,659 & 5,349 & 1,829 & 22,278 \\ 
David & 4,739 & {\bf5,375 (+10\%)} & 4,526 & 4,858 & 1,657 & 21,155 \\ 
John & 4,590 & 3,979 & {\bf5,154 (+9\%)} & 4,687 & 1,598 & 20,008 \\ 
Chris & 3,971 & 3,431 & 3,659 & {\bf4,791 (+9\%)} & 1,410 & 17,262 \\ 
James & 2,349 & 1,980 & 2,329 & 2,556 & {\bf1,287 (+30\%)} & 10,501 \\ 
\hline
Total & 21,598 & 19,257 & 20,327 & 22,241 & 7,781 & \\ 
\hline
\end{tabular}
}
\end{center}

\caption{Results for the SNE on popular names in the US on Twitter and Google+. The effect sizes are positive for all five names showing the SNE ($p-values < 0.001$).}
\label{table:SNE}
\vspace{-5mm}
\end{table*}

\begin{table*}[tbh!]

\begin {center}
\texttt{Twitter}\\ 
{
\begin {tabular} { c c c c c c c c}
\hline
Race & \specialcell{\# of \\ last names} & \specialcell{Could be \\ tested} & \specialcell{Median\\SNE} & \specialcell{Sig. \\ positive} & \specialcell{Sig. \\ negative} & \specialcell{Non-sig. \\ positive} & \specialcell{Non-sig. \\ negative} \\ 
\hline
White & 35 & 1225 & 45\% & 986 & 0 & 237 & 2\\
Asian & 394 & 1193 & 59\% & 433 & 1 & 645 & 114\\
Hispanic & 341& 1073 & 86\% & 350 & 0 & 541 & 146\\
Native American & 72 & 345 & 100\% & 78 & 0 & 221 & 46\\
African-American & 64 & 263 & 100\% & 56 & 0 & 241 & 66\\
\end{tabular}
}
\end{center}

\begin {center}
\texttt{Google+}\\ 
{
\begin {tabular} { c c c c c c c c}
\hline
Race & \specialcell{\# of \\ last names} & \specialcell{Could be \\ tested} & \specialcell{Median\\SNE} & \specialcell{Sig. \\ positive} & \specialcell{Sig. \\ negative} & \specialcell{Non-sig. \\ positive} & \specialcell{Non-sig. \\ negative} \\ 
\hline
White            & 35  & 80  & 43\%  & 0  & 0 & 79  & 1  \\
Asian            & 394 & 608 & 47\%  & 61 & 0 & 494 & 53  \\
Hispanic         & 341 & 95  & 60\%  & 0  & 0 & 95  & 0  \\
Native American  & 72  & 1   & -17\% & 0  & 0 & 0   & 1  \\
African-American & 64  & 16  & 18\%  & 0  & 0 & 16  & 0  \\
\end{tabular}
}
\end{center}

\caption{The SNE test for users with a race-specific last name. The ``\# of last names'' indicates the number of race-specific last names found. ``Could be tested'' is the number of first name pairs where each first name had a non-zero count for the race-specific last name. There is a large number of statistically significant positive effects, and only a single first name pair with a significant \emph{negative} effect.}
\label{table:race}
\vspace*{-3mm}
\vspace*{-3mm}
\end{table*}

\vspace{-1mm}
\subsection{SNE and Social Tie Strength}
\vspace{-2mm}
%\todo[Gabriel]{Include tie strength results for Google+}

We also investigated the correlation of the SNE and the strength of the tie between users. Concretely,
are users' strong ties more affected by the SNE than their weak ties? Again, we limited
this analysis to users from the US and the mentioned popular names of American users on Twitter. 

We eliminated all the super-users to better capture the strong and weak ties among normal users.
For all normal users, for the link from user $A$ to $B$, we looked at the Jaccard similarity of the friends of 
$A$ and $B$ as a measure of the strength of the tie. Then, we considered half of the links
with the lower strength as weak links and the other half as strong ties ($threshold = 0.008$).
First, we tested the SNE by only considering weak ties, and then by only considering strong ties. In both cases,
the SNE was statistically significant. For weak links the preference ranged
from 13\% to 17\% and for strong ties from 10\% to 13\%, and for all the five names the SNE was
slightly stronger for the weak ties.
Our results suggest that people are more affected by SNE when they are establishing a weak link.
This is in contrast with an earlier study that has found the NLE only affects people's
 important decisions, such as choosing a job or place to live, and not the more trivial decisions
like favorite animals or foods~\cite{hodsonolson05pspb}. This observation was explained by an 
earlier finding that the NLE is a type of implicit egotism and implicit egotism is boosted under stress~\cite{Johnson1989}.
However, we do not find evidence to support this finding in Twitter. Though, the results on Google+ are not consistent with these findings and the preference for weak ties ranged from 7-33\% and for strong ties 12-45\%. %This means on Google+ strong ties are more under the influence of the SNE compared to the weak links.
 Further investigation of differences between Twitter and Google+ is needed to figure out the root of the mentioned inconsistency.

\vspace{-2mm}
\subsection{SNE and Number of Friends}
\vspace{-2mm}
%\todo[Gabriel]{Verify y-scale}

Finally, we examined the correlation between the SNE and the number of friends (followees) of users.
The aim is to see if the SNE differs for users with more compared to
users with less friends. Similar to before, we considered only
users from the US and the mentioned popular names on Twitter.
Then, we grouped users based on their number of friends logarithmically, up to 64 friends and a group for users with more than 64 friends.
The resulting groups are fairly balanced, with the smallest group (one friend) containing 
8\% of considered users and the largest group (between 16 and 32 friends) 20\% of them. We also use the same group sizes for Google+.

\begin{figure} [bth!]
\centering
\vspace{-3mm}
\hspace*{-6mm}
{\includegraphics[width=.60\textwidth]{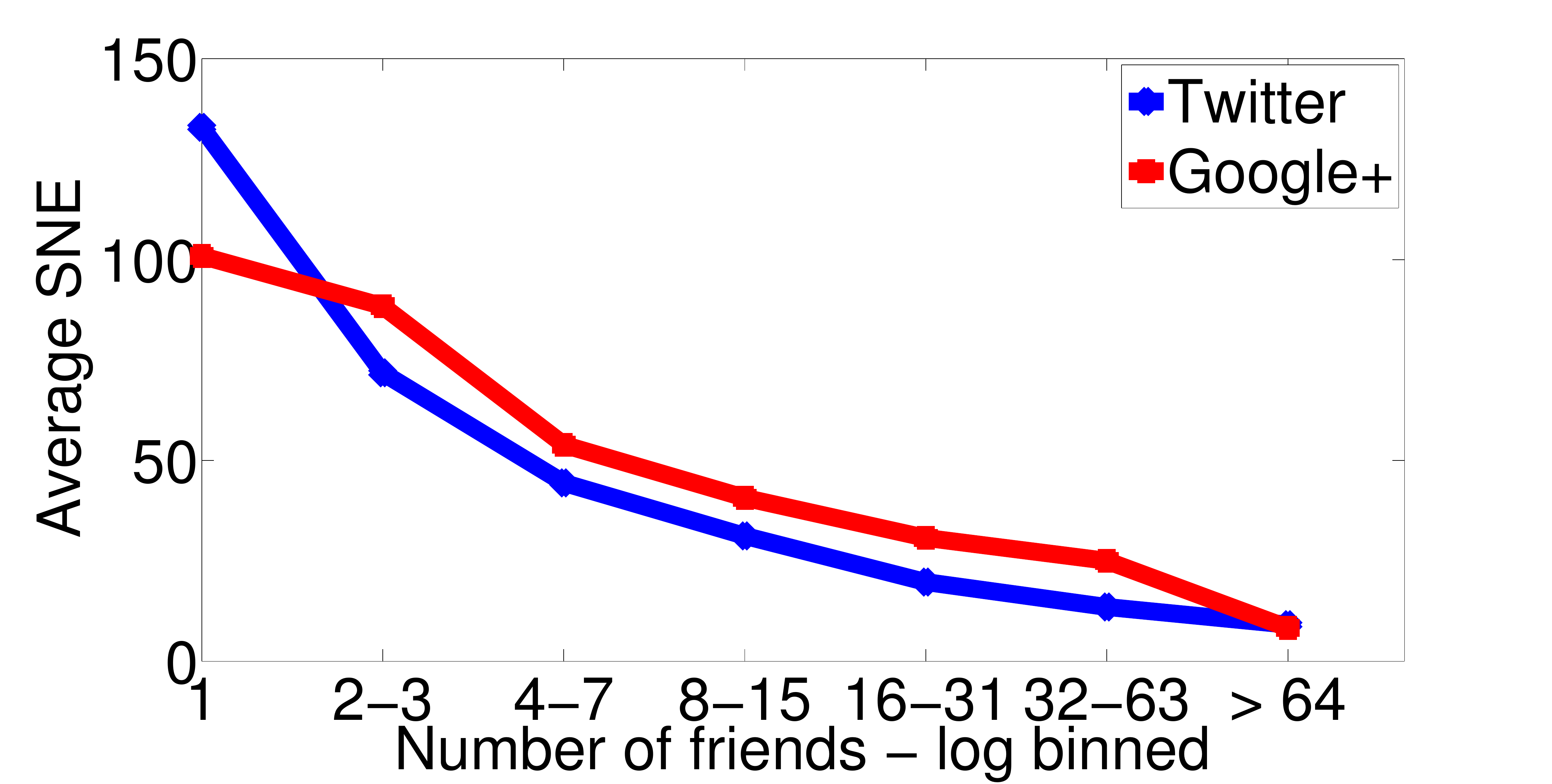}}
\vspace{-3mm}
\caption{The average SNE of users grouped by the number of friends.}
\label{fig:deg_SNE}
\vspace{-3mm}
\end{figure}

We tested the SNE in each of the groups by only considering the links going out from users
of that group and then taking the average of the SNE for the five considered names. 
Figure~\ref{fig:deg_SNE} shows that there is a noticeable reverse correlation between
the number of friends and the SNE. Users with fewer friends are more likely to follow other
users with the same name compared to the users with a higher number of friends.

\end{document}